\documentclass{article}

\usepackage{microtype}
\usepackage{graphicx}
\usepackage{subfigure}
\usepackage{booktabs} 

\usepackage{hyperref}

\usepackage[accepted]{icml2025}

\usepackage{amsmath}
\usepackage{amssymb}
\usepackage{mathtools}
\usepackage{amsthm}

\usepackage[capitalize,noabbrev]{cleveref}

\theoremstyle{plain}

\theoremstyle{definition}

\theoremstyle{remark}

\usepackage[textsize=tiny]{todonotes}

\icmltitlerunning{Submission and Formatting Instructions for ICML 2025 GenBio Workshop}

\begin{document}

\twocolumn[
\icmltitle{Fast and Scalable Gene Embedding Search:\\
           A Comparative Study of FAISS and ScaNN}



\icmlsetsymbol{equal}{*}

\begin{icmlauthorlist}
\icmlauthor{Mohammad Saleh Refahi}{yyy}
\icmlauthor{Gavin Hearne}{yyy}
\icmlauthor{Harrison Muller}{yyy}
\icmlauthor{Kieran Lynch}{yyy}
\icmlauthor{Bahrad A. Sokhansanj}{yyy}
\icmlauthor{James R. Brown}{yyy}
\icmlauthor{Gail Rosen}{yyy}

\end{icmlauthorlist}

\icmlaffiliation{yyy}{Department of Electrical and Computer Engineering, Drexel University, Philadelphia, USA}

\icmlcorrespondingauthor{Gail Rosen}{glr26@drexel.edu}
\icmlcorrespondingauthor{Mohammad Saleh Refahi}{sr3622@drexel.edu}

\icmlkeywords{Machine Learning, ICML}

\vskip 0.3in
]



\printAffiliationsAndNotice{\icmlEqualContribution} 

\begin{abstract}
The exponential growth of DNA sequencing data has outpaced traditional heuristic-based methods, which struggle to scale effectively. Efficient computational approaches are urgently needed to support large-scale similarity search, a foundational task in bioinformatics for detecting homology, functional similarity, and novelty among genomic and proteomic sequences. Although tools like BLAST have been widely used and remain effective in many scenarios, they suffer from limitations such as high computational cost and poor performance on divergent sequences.

In this work, we explore embedding-based similarity search methods that learn latent representations capturing deeper structural and functional patterns beyond raw sequence alignment. We systematically evaluate two state-of-the-art vector search libraries—FAISS and ScaNN—on biologically meaningful gene embeddings. Unlike prior studies, our analysis focuses on bioinformatics-specific embeddings and benchmarks their utility for detecting novel sequences, including those from uncharacterized taxa or genes lacking known homologs. Our results highlight both computational advantages (in memory and runtime efficiency) and improved retrieval quality, offering a promising alternative to traditional alignment-heavy tools.
\end{abstract}

\section{Introduction}

Traditional alignment-based methods such as BLAST\cite{altschul1990basic} and MMseqs2\cite{steinegger2017mmseqs2} have long served as the backbone of sequence comparison. While effective in many cases, these approaches are limited by their reliance on exact or near-exact matches, making them less suitable for detecting distant homology or structural variation—particularly in short or noisy sequences. Moreover, their computational cost scales poorly with data volume, posing significant challenges for large-scale genomic analysis .

These limitations are especially pronounced in metagenomics, where datasets are highly fragmented, taxonomically diverse, and often contain a substantial proportion of sequences with no close reference. But similar challenges extend beyond metagenomics—affecting areas such as epigenomics , gene regulation studies\cite{mathebela2022influence,safaei2025caspase}, and disease-related variant discovery, where subtle patterns or regulatory signals may be obscured by noise or evolutionary divergence\cite{liu2024correlation}.

\begin{figure*}[h]
    \centering
    \includegraphics[width=0.85\textwidth]{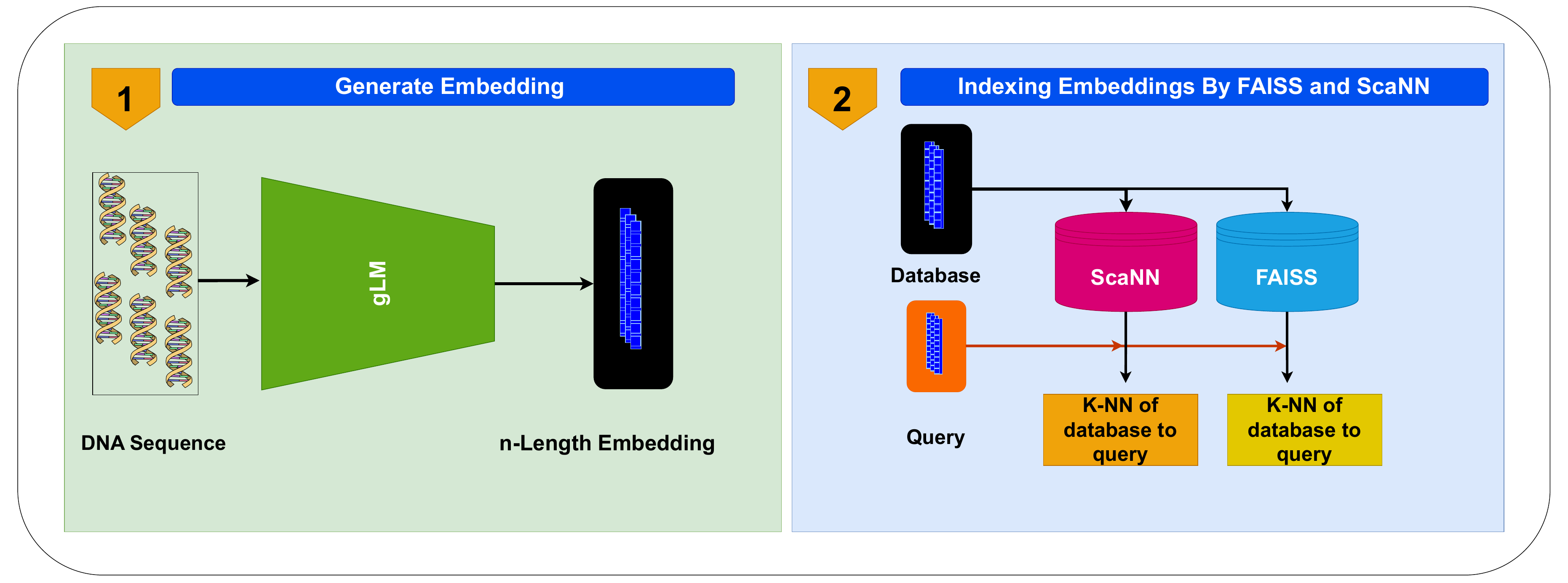}
    \caption{Overview of the embedding-based retrieval pipeline. Sequences are first encoded into dense vector representations using a pretrained language model. These embeddings are then indexed using similarity search methods with FAISS or ScaNN to enable fast and efficient nearest-neighbor retrieval.}
    \label{fig:pip}
\end{figure*}

Inspired by advances in natural language processing, genomic language models now leverage representation learning to extract biologically meaningful embeddings from raw DNA sequences. These embeddings capture structural, functional, and evolutionary signals in dense vector spaces, allowing comparisons that go beyond exact sequence alignment. Recent models—such as DNABERT~\cite{ji2021dnabert,zhou2023dnabert}, MetaBERTa~\cite{refahi2023leveraging}, Nucleotide Transformer~\cite{dalla2024nucleotide}, HyenaDNA~\cite{nguyen2024hyenadna}, and Caduceus~\cite{schiff2024caduceus}—have demonstrated strong performance on a range of genomics and metagenomics tasks by learning contextualized nucleotide representations.

To make these embeddings useful in practice—particularly for tasks like novelty detection, taxonomic classification, or gene retrieval—we need fast and scalable similarity search infrastructure. Unlike traditional models that rely on token-level prediction, these downstream applications require efficient nearest neighbor retrieval across large embedding databases. Approximate nearest neighbor (ANN) methods offer a promising solution, but their performance in biological settings remains underexplored. In this work, we systematically evaluate two state-of-the-art ANN libraries, FAISS~\cite{johnson2019billion, douze2024faiss} and ScaNN~\cite{guo2020accelerating}, assessing their speed, accuracy, and utility for real-world metagenomic retrieval.

Beyond comparing their default configurations, we conduct detailed parameter tuning for both tools to evaluate how internal settings—such as index type, quantization strategy, distance metric, and search depth—affect performance. We report their impact on runtime, memory usage, and retrieval quality. Our results provide actionable insights for practitioners seeking scalable and accurate search infrastructure for biological embeddings.

\section{Background}
Deep learning methods for biological sequences have made it possible to learn embeddings that capture functional and evolutionary patterns, enabling similarity search via vector comparisons \cite{schutze2022nearest}. These approaches allow scalable analysis by avoiding expensive alignment computations.

\textbf{FAISS}, developed by Facebook AI Research, supports various ANN indexing structures including inverted files (IVF), product quantization (PQ), and HNSW graphs. It supports both L2 and inner product distance metrics and runs on CPU and GPU. FAISS is highly customizable, offering flexibility for tuning indexing and search parameters to match dataset scale and hardware constraints \cite{johnson2019billion, douze2024faiss, jegou2011PQ}. It has been successfully applied in multilingual NLP and image retrieval, and recently in bioinformatics tasks like immune receptor clustering \cite{10.1093/bioinformatics/btab446}.

\textbf{ScaNN}, developed by Google Research, introduces enhancements such as score-aware and anisotropic quantization losses and asymmetric distance computation, improving precision and recall at scale \cite{guo2020accelerating, xiong2020approximate, borgeaud2022improving}. It integrates well with TensorFlow pipelines and supports SIMD acceleration for high-speed computation. ScaNN has shown effectiveness in billion-scale tasks like entity linking \cite{fitzgerald2022moleman}, and its simplicity makes it a good candidate for bioinformatics applications, especially when ease of deployment is a priority.

Despite these advantages, both systems require careful parameter selection to achieve optimal performance. This work explores how such parameters impact their behavior in a biological setting, especially for tasks involving short gene embeddings from diverse taxa.

\begin{figure*}[h]
    \centering
    \includegraphics[width=0.47\textwidth]{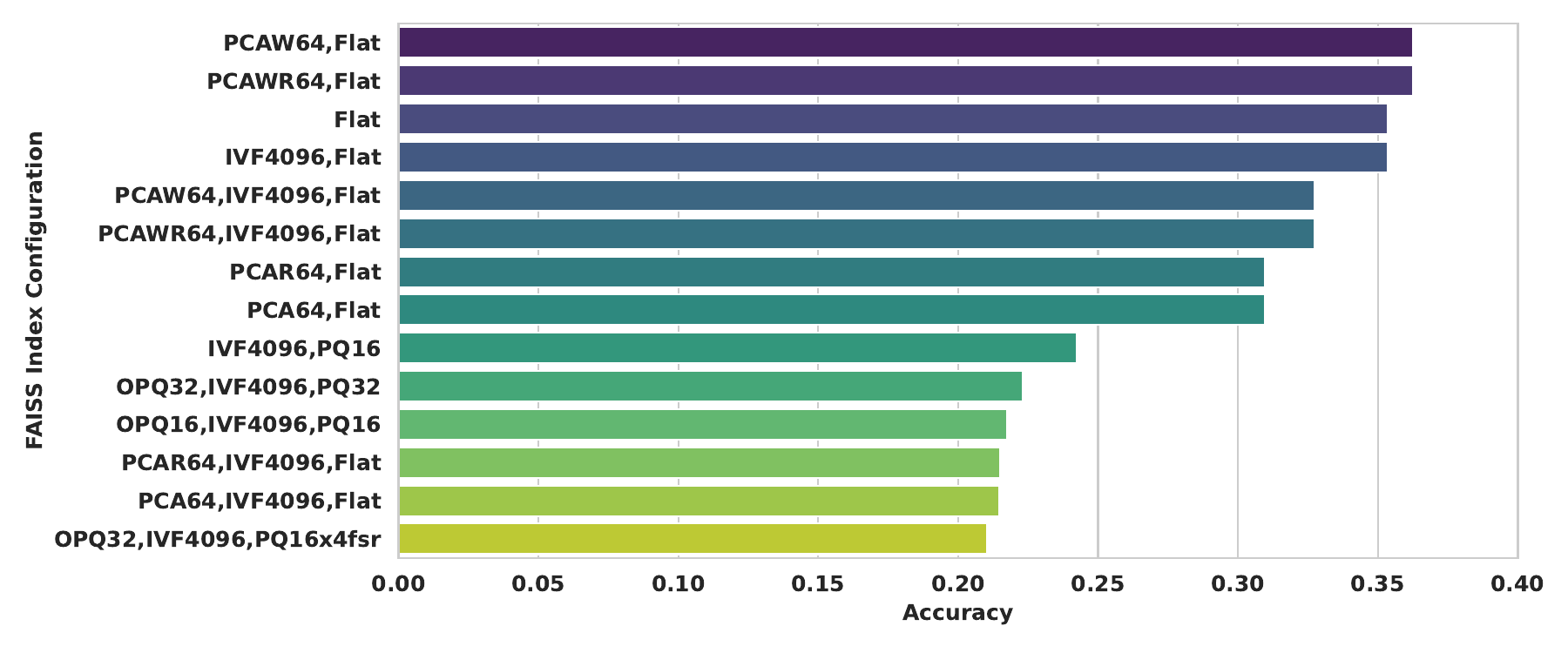}
    \hfill
    \includegraphics[width=0.5\textwidth]{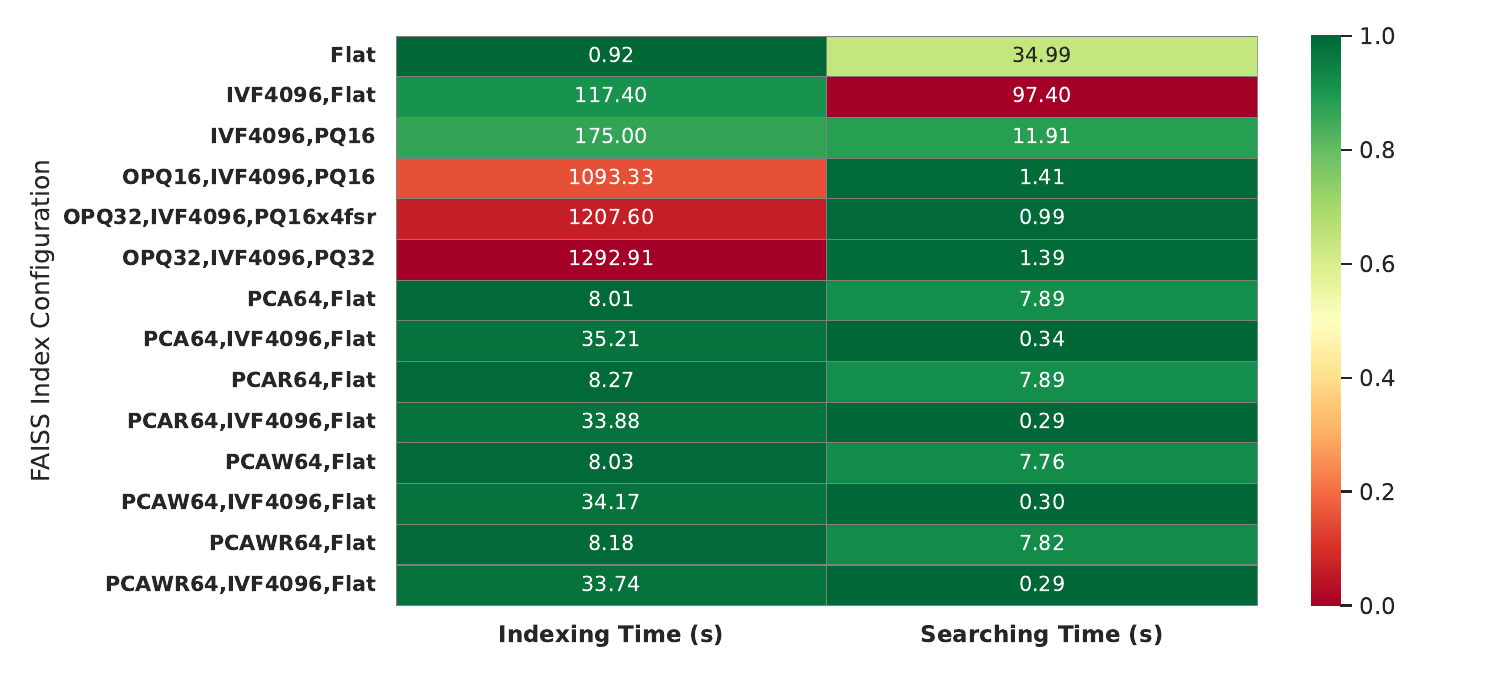}
    \caption{Performance of FAISS index configurations. \textbf{Left:} Classification accuracy across indexing strategies. PCA-enhanced Flat indexes (e.g., PCAW64,Flat) yield the best accuracy, outperforming the pure Flat index. In contrast, quantization-based approaches like OPQ and PQ reduce accuracy due to lossy compression. \textbf{Right:} Latency heatmap. IVF-based combinations with OPQ/PQ deliver the fastest search times, making them ideal for high-throughput querying, though they come at the cost of higher indexing time and lower accuracy. PCA+Flat offers a balanced trade-off, combining strong accuracy with reasonably fast indexing and querying.}
    \label{fig:faiss_combined_simple}
\end{figure*}

\section{Materials and Methods}

We benchmark FAISS and ScaNN for gene embedding search tasks in metagenomics. Both tools are evaluated using dense vector representations of microbial gene fragments. Our objective is to assess their scalability and effectiveness for fast similarity search, with an emphasis on parameter tuning and performance across in-domain and out-of-domain sequences.

\subsection*{Dataset}
We use the \textit{Scorpio-Gene-Taxa (short fragment version)}~\cite{refahi2025enhancing} dataset, which consists of 400 bp DNA fragments sampled from 497 housekeeping genes across 2,046 bacterial and archaeal genera. This dataset is designed for evaluating generalization in embedding-based metagenomic analysis.

While the full dataset contains a large training set, we use the {test set}—comprising 165{,}615 sequences—as our {search database} for all retrieval tasks. This choice is motivated by computational efficiency: using the smaller test set enables full pairwise distance computation between all queries and the entire database, which is essential for retrieval, novelty detection, and clustering analyses.

We define two query sets for evaluation:
\begin{itemize}
    \item \textbf{In-domain set:} 7{,}000 sequences from species seen during training but with non-overlapping gene fragments.
    \item \textbf{Out-domain set:} 7{,}000 sequences from phyla not present in the training set, used to assess generalization to unseen taxa.
\end{itemize}

\subsection*{Generating Embeddings}

To generate sequence embeddings, we use the \textit{MetaBERTA-BigBird}\cite{refahi2025enhancing,refahi2023leveraging} model, a transformer-based language model trained on microbial gene sequences. Both the in-domain and taxa-out query sets are encoded into dense vectors using this model. Each sequence is represented as a 1{,}024-dimensional embedding, which is used as input for the similarity search. Embedding distances between the query sets and the database are computed using FAISS and ScaNN. These distances are analyzed to infer novelty at the phylum level and to support unsupervised clustering, enabling downstream biological interpretation of embedding quality and retrieval performance.

\vspace{0.5em}
\subsection*{FAISS Configuration}  
We explore multiple FAISS indexing strategies via the \texttt{indexfactory()} interface:

\begin{itemize}
    \item \textbf{Flat Index}: Brute-force exact search, storing all vectors without compression.
    \item \textbf{IVF (Inverted File Index)}: Partitions the vector space using a coarse quantizer and probes a subset during search.
    \item \textbf{PQ (Product Quantization)}: Compresses vectors into sub-quantized codes to reduce memory and speed up distance computation.
\end{itemize}

We also test preprocessing strategies:
\begin{itemize}
    \item \textbf{PCA}: Dimensionality reduction to speed up indexing.
    \item \textbf{OPQ}: Rotational transformation that improves PQ precision.
\end{itemize}

For IVF, we sweep the number of probes from 100 to 1000 and evaluate accuracy, memory usage, and runtime for each configuration.

\vspace{0.5em}
\subsection*{ScaNN Configuration}  
We evaluate ScaNN’s vector search performance by manually configuring its index and search pipeline, consisting of a required scoring step and optional re-scoring and partitioning:

\begin{itemize}
    \item \textbf{Partitioning}: Divides the entire vector dataset into clusters during an offline training step to reduce the number of candidate vectors considered at query time.
    
    \item \textbf{Brute Force}: Computes exact distances or inner products between the query and all vectors (or all vectors in selected partitions). Datapoint integers may also be quantized (while maintaining brute force scoring).
    
    \item \textbf{Asymmetric Hashing (AH)}: A fast approximate scoring method where database vectors are quantized using product quantization, and queries remain in full precision. 
    
    \item \textbf{Reordering}: An optional refinement stage in which the top-$k$ candidates from the initial scoring step are re-ranked using full-precision exact inner products. 

\end{itemize}

We explore configurations that prioritize either accuracy or speed by adjusting the recall target and partition size.

\begin{figure*}[h]
    \centering
    \includegraphics[width=0.49\textwidth]{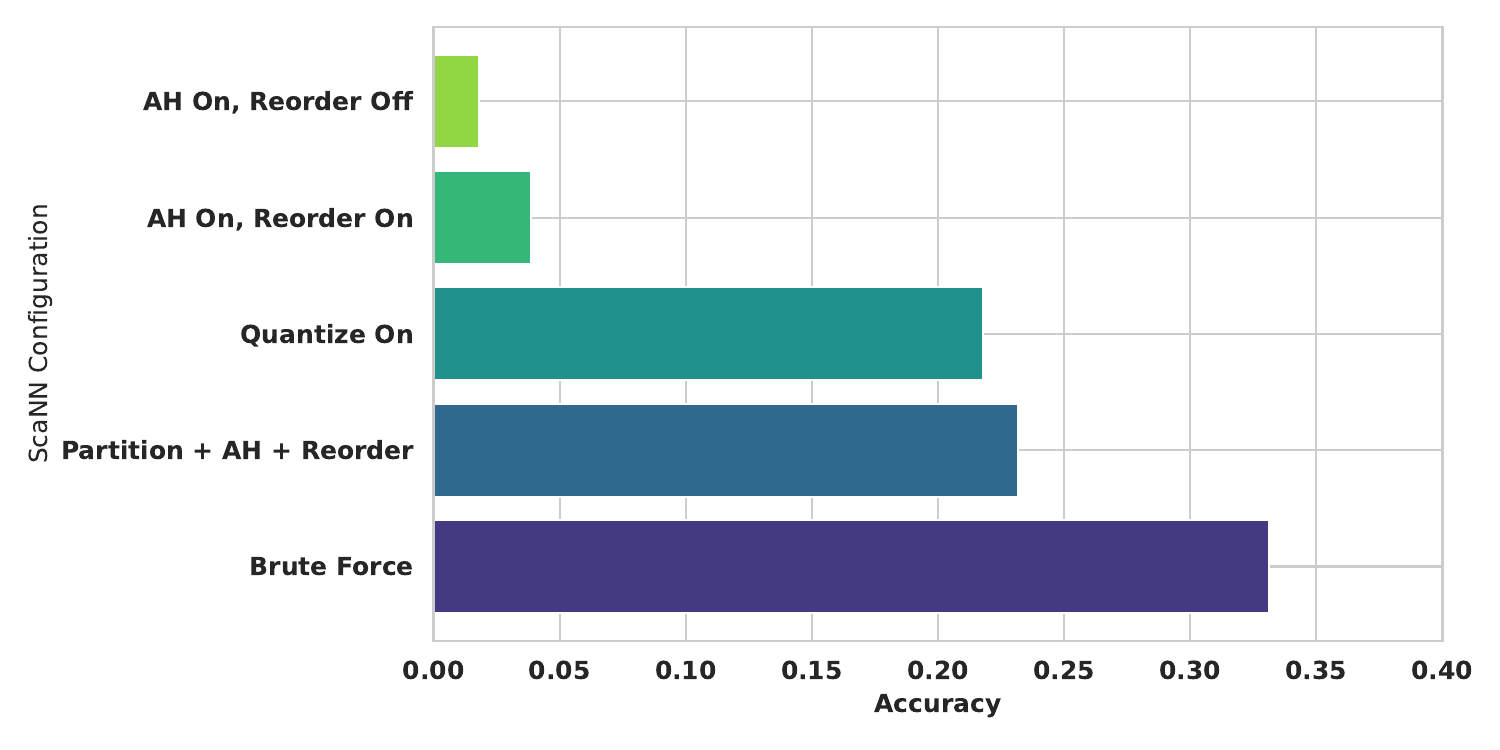}
    \hfill
    \includegraphics[width=0.49\textwidth]{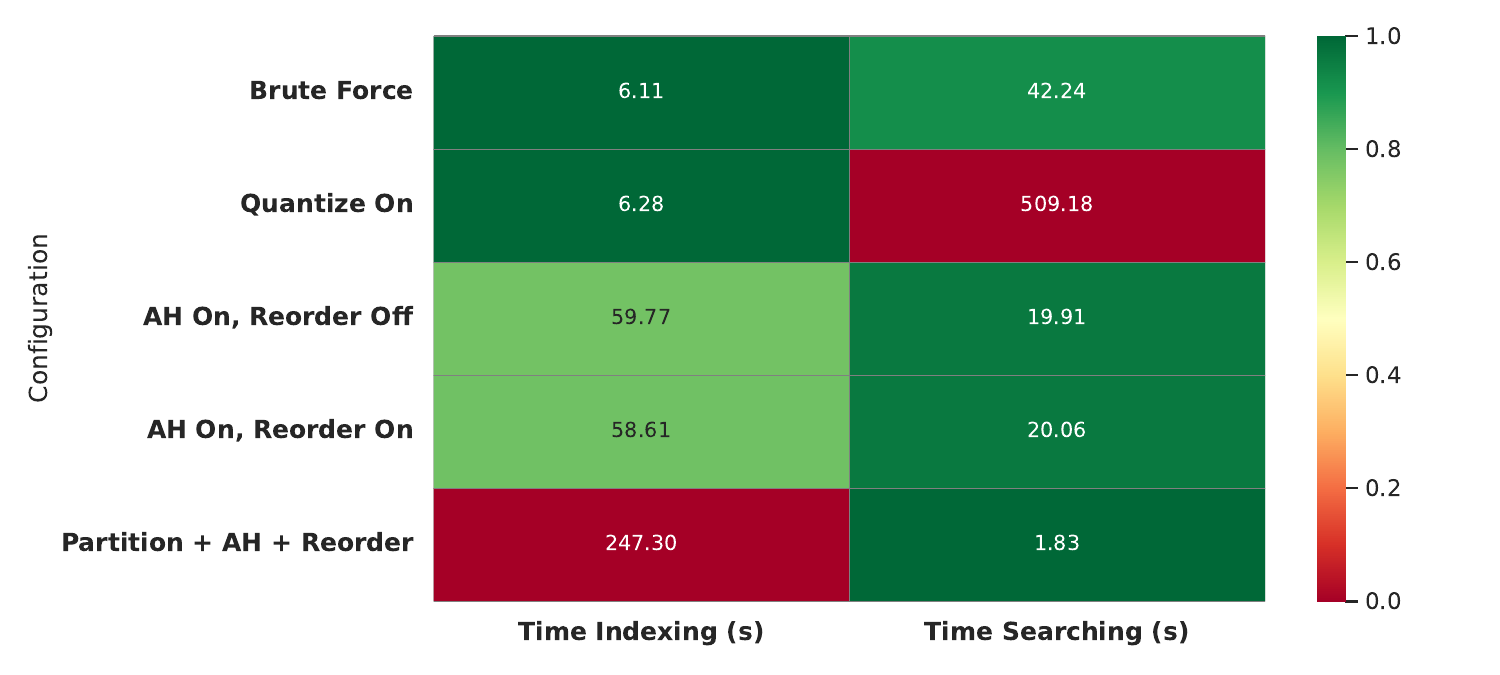}
    \caption{ScaNN parameter evaluation.\textbf{Left:} Classification accuracy of each configuration, with error bars showing variance across trials. Brute force search achieves the highest accuracy, while asymmetric hashing without reordering performs poorly. \textbf{Right} Heatmap of average indexing and query latency across different configurations. Green indicates lower latency. Partitioning with asymmetric hashing and reordering achieves the fastest performance. }
    \label{fig:scann_combined}
\end{figure*}

\begin{figure*}[h]
    \centering
    \includegraphics[width=\textwidth]{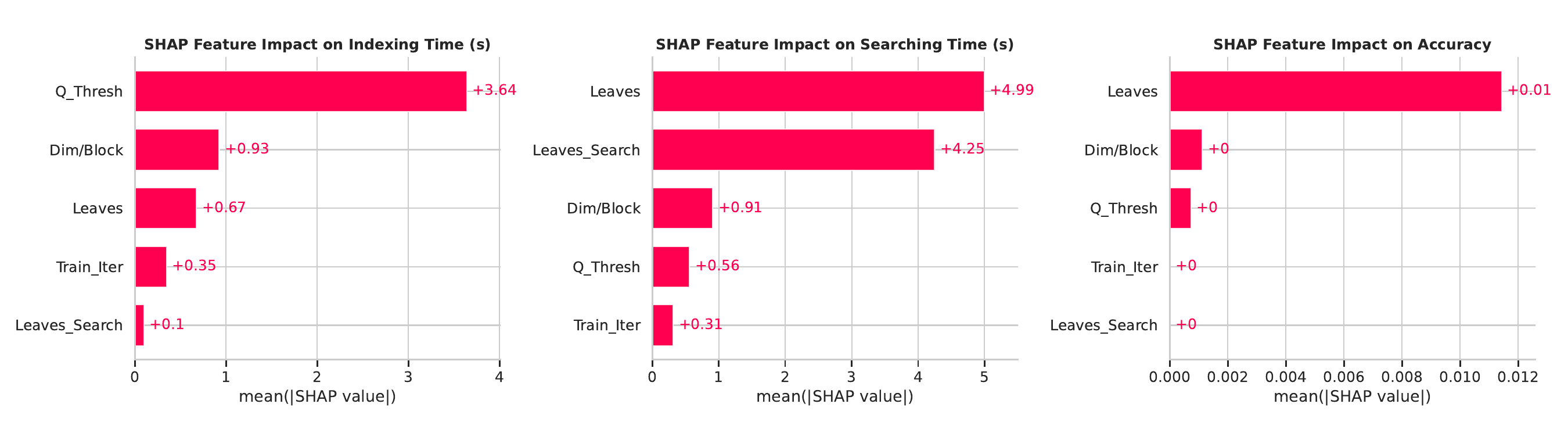}
    \caption{SHAP summary plots showing the average impact of each parameter on build time (left), query time (center), and classification accuracy (right) for ScaNN. Query time is most influenced by the number of tree Leaves and Leaves\_Search, while build time is governed by Q\_Thresh and Dim/Block. Accuracy is largely insensitive to these parameters in this configuration space.}
    \label{fig:scann_shap_summary}
\end{figure*}

\section{Results}

\subsection*{FAISS Parameter Sensitivity}

To explore the trade-offs between accuracy and performance, we conducted an extensive sweep of FAISS indexing methods using the \texttt{indexfactory()} interface. We evaluated configurations including Flat, IVF, and Product Quantization (PQ), with and without preprocessing methods such as PCA and OPQ.

Figure~\ref{fig:faiss_combined_simple} presents the classification accuracy of each configuration. The highest top-1 accuracy was achieved by PCA-enhanced Flat configurations—specifically \texttt{PCAW64,Flat} and \texttt{PCAWR64,Flat}—both reaching an accuracy of {0.362}. These configurations slightly outperformed the pure \texttt{Flat} index and \texttt{IVF4096,Flat} (both at {0.353}). In contrast, heavily compressed configurations such as \texttt{IVF4096,PQ16} and \texttt{OPQ32,IVF4096,PQ16x4fsr} exhibited substantial accuracy drops, with values as low as {0.210}. These results suggest that aggressive quantization compromises the fidelity of similarity matching, likely due to coarse-grained distance representation in the embedding space. For downstream tasks such as gene annotation or novel gene detection, this reduction in precision may hinder biological interpretability.

However, accuracy alone does not determine the utility of a method in large-scale retrieval systems. The latency heatmap in Figure~\ref{fig:faiss_combined_simple} visualizes the trade-offs in both indexing and querying time. Notably, OPQ and PQ methods—particularly those combined with IVF—achieve the fastest query times, with \texttt{OPQ32,IVF4096,PQ16x4fsr} reaching a search time of just {0.99s}, though requiring an indexing time of over {1207s}. These gains come at the cost of substantial indexing overhead, with the slowest configurations (e.g., \texttt{OPQ32,IVF4096,PQ32}) requiring up to {1293s} to build the index.

In contrast, \texttt{Flat} and PCA-enhanced Flat indexes such as \texttt{PCAW64,Flat} and \texttt{PCAWR64,Flat} exhibit low indexing time (around {8.0s}) and competitive query speeds (around {7.8s}), while maintaining the highest accuracy. \texttt{IVF4096,Flat} also offers strong accuracy ({0.353}), but suffers from significantly higher search latency ({97.4s}).

These results highlight that FAISS parameter tuning must align with task-specific constraints. For applications where accuracy is paramount—such as taxonomy prediction or gene similarity analysis—PCA-enhanced Flat indexing is preferred. For applications requiring rapid large-scale retrieval with relaxed accuracy demands, OPQ or PQ combined with IVF provide an efficient alternative.

\begin{figure*}[h]
    \centering
    \includegraphics[width=0.47\textwidth]{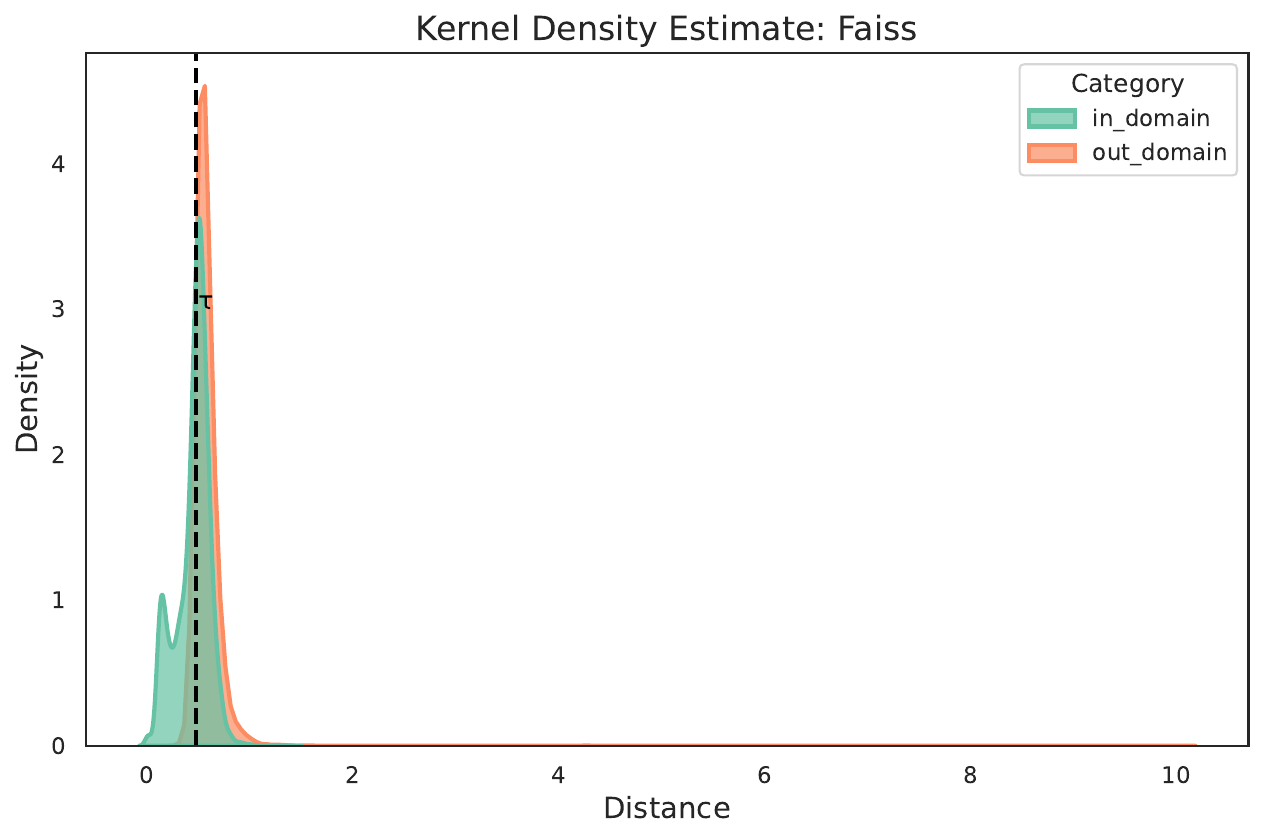}
    \hfill
    \includegraphics[width=0.47\textwidth]{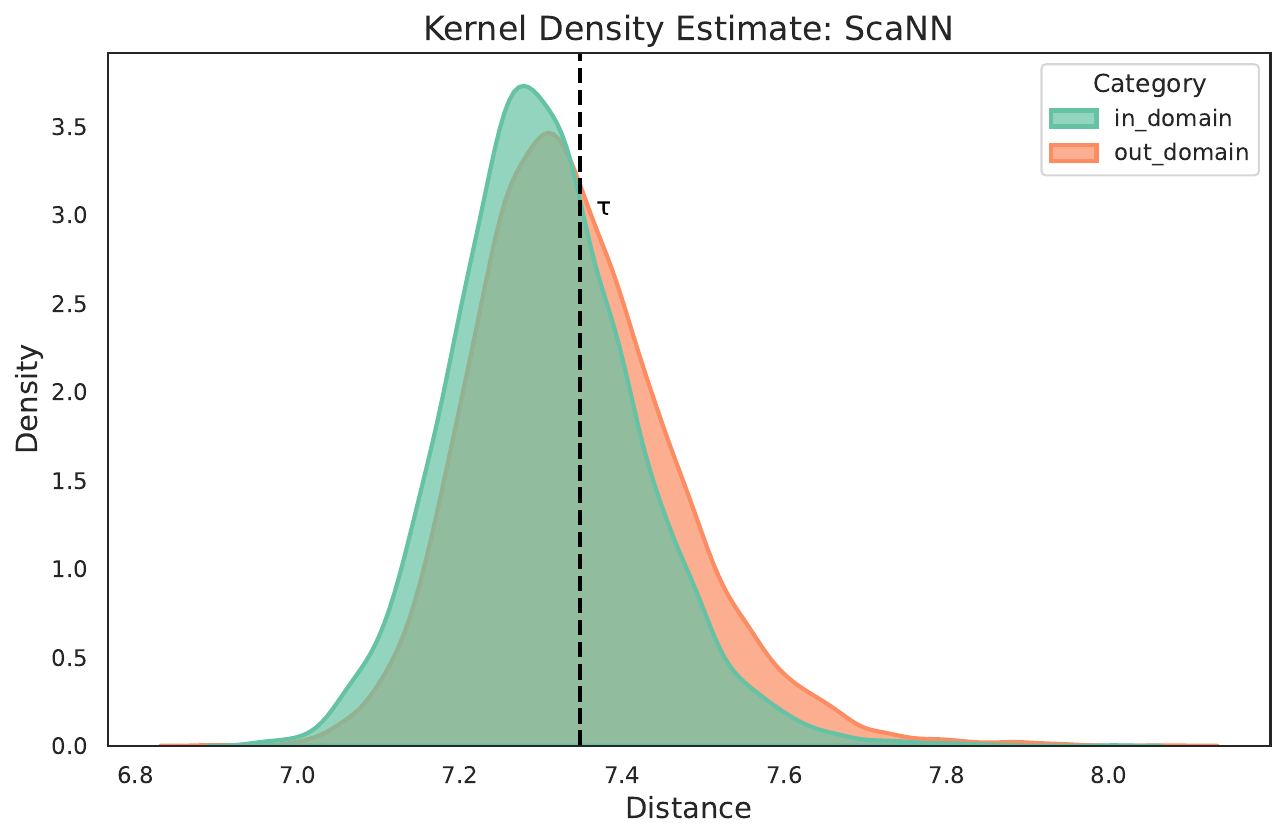}
    \caption{
    Kernel density estimation (KDE) plots of embedding distances for in-domain and out-of-domain queries using FAISS (left) and ScaNN (right). While neither method achieves a clearly separated distribution, FAISS shows a slightly broader and more interpretable distance range, with a stronger low-distance peak for in-domain queries. This suggests marginally better utility for novelty detection compared to ScaNN.
    }
    \label{fig:faiss_scann_kde}
\end{figure*}

\subsection*{ScaNN Configuration Sensitivity}

For ScaNN, we evaluated five settings combining brute-force scoring, asymmetric hashing (AH), and optional partitioning and reordering (Figure~\ref{fig:scann_combined}). 

Brute-force search with full-precision vectors serves as the upper bound for accuracy, achieving consistent top-1 accuracy of 0.33. However, it incurs substantial query latency ($\sim$42 seconds per run), limiting scalability. Enabling quantization during brute-force scoring severely degrades both performance and accuracy due to the relatively small nature of our dataset, increasing query time by 10x and dropping accuracy to 0.22.

In contrast, pure AH-based scoring configurations (with no partitioning) offer significantly faster query speeds ($\sim$20s), but with a steep drop in accuracy: only 1.8\% without reordering and 3.9\% with reordering. This implies that quantization alone is insufficient for effective inner product approximation in this scenario. Combining AH with both reordering and partitioning yields the best runtime ($\sim$1.8s query latency) while maintaining moderate accuracy ($\sim$0.25), though results are more variable across runs—likely due to the stochastic nature of clustering and quantization.

As expected, brute-force methods maximize accuracy but are impractical at scale, while AH-based configurations are suitable for real-time querying if some accuracy loss is tolerable. The combined use of partitioning, quantization, and reordering appears to offer the best compromise for large-scale applications requiring both speed and acceptable recall, as is common in genomic and metagenomic tasks where samples may be particularly large.

\subsection*{ScaNN Parameter Tuning}

To identify which parameters most influence ScaNN's behavior, we performed SHAP \cite{lundberg2017unified} analysis over a sweep of configurations, measuring their impact on indexing time, query time, and classification accuracy (Figure~\ref{fig:scann_shap_summary}). The results show that indexing time is primarily affected by \texttt{Q\_Thresh} and \texttt{Dim/Block}, which govern the precision and granularity of the quantization process. In contrast, query time is dominated by \texttt{Leaves} and \texttt{Leaves\_Search}, which control the number of partitions searched and the scope of candidate retrieval. Interestingly, SHAP scores for accuracy were near zero across all parameters, indicating that within the tested range, ScaNN’s performance on classification is relatively insensitive to parameter tuning. These findings suggest that ScaNN tuning is most effective for optimizing runtime and memory efficiency, rather than accuracy. For example, using fewer partitions or search leaves can reduce latency significantly without compromising retrieval quality. Brute-force search still yields the highest accuracy, but tuned configurations—particularly those combining partitioning, asymmetric hashing, and reordering—offer favorable trade-offs between speed and performance.

\subsection*{Similarity Search Performance Comparison}

To benchmark similarity search performance, we compare FAISS, ScaNN, and MMseqs2 across accuracy, runtime, and memory usage. We selected the best-performing configurations for FAISS and ScaNN based on top-1 retrieval accuracy from a parameter sweep.

FAISS with the \texttt{PCA64,Flat} configuration achieved the highest accuracy (36.2\%). However, FAISS with \texttt{PCAWR64,IVF4096,Flat} offers a significantly faster search time (0.3 s vs.\ 7.7 s)—about 25× faster—while maintaining competitive accuracy (32.7\%). This makes it a strong practical choice when inference time is critical.

ScaNN, while slightly less accurate (31\%), demonstrated efficient search performance (2.1 s) and low memory usage. MMseqs2~\cite{steinegger2017mmseqs2}, although extremely lightweight in indexing memory (99.5 MB), struggled with both low accuracy (1.8\%) and slow search time (25.7 s), likely due to the limitations of alignment-based methods on short 400 bp fragments.

As summarized in Table~\ref{tab:faiss_scann_summary}, embedding-based methods like FAISS and ScaNN consistently outperform alignment-based approaches for short-fragment similarity search in terms of both speed and accuracy.

Figures~\ref{fig:faiss_combined_simple} and~\ref{fig:scann_combined} present a side-by-side comparison of performance across indexing configurations. FAISS’s \texttt{PCA64,IMI2x10,Flat} configuration achieves the fastest query time at 0.12 seconds. Additionally, the \texttt{IMI2x10,Flat} setup maintains a low query time of approximately 0.76 seconds—about 2 times better than ScaNN’s best-performing configuration. ScaNN's optimal setup, which combines partitioning, asymmetric hashing (AH), and reordering, achieves a query latency of 1.83 seconds. These results underscore FAISS’s superior efficiency in large-scale embedding retrieval tasks.

\begin{table}[htbp]
\centering
\caption{Summary comparison between FAISS, ScaNN, and MMseqs2 on accuracy, runtime, and memory usage.}
\label{tab:faiss_scann_summary}
\resizebox{\columnwidth}{!}{%
\begin{tabular}{l|c|cc|cc}
\toprule
\textbf{Model} & \textbf{Accuracy (↑)} & \multicolumn{2}{c|}{\textbf{Time (s) (↓)}} & \multicolumn{2}{c}{\textbf{Memory (MB) (↓)}} \\
\cmidrule(lr){3-4} \cmidrule(lr){5-6}
 &  & \textbf{Index} & \textbf{Search} & \textbf{Index} & \textbf{Search} \\
\midrule
FAISS*        & \textbf{0.362} & 8.0 & 7.7 & 647.0 & 56.0 \\
FAISS†        & {0.327} & 33.7 & \textbf{0.3} & 647.0 & 56.0 \\
ScaNN         & 0.310 & 240.9 & 2.1 & 646.9 & \textbf{55.8} \\
MMseqs2       & 0.018 & \textbf{5.8} & 25.7 & \textbf{99.5} & 209.2 \\
\bottomrule
\end{tabular}%
}
\vspace{1mm}
\begin{minipage}{0.95\columnwidth}
\footnotesize
\textbf{*} FAISS using \texttt{PCA64,Flat} — highest accuracy. \\
\textbf{†} FAISS using \texttt{PCAWR64,IVF4096,Flat} — best speed–accuracy tradeoff (25× faster).
\end{minipage}
\end{table}

Figure~\ref{fig:faiss_scann_kde} presents the distance distributions between query embeddings and the database for both FAISS and ScaNN. Although neither method produces a sharply separated distribution between in-domain and out-of-domain queries, FAISS shows a slightly better contrast. Its KDE plot reveals a broader dynamic range and a more pronounced peak at lower distances for in-domain sequences. This mild separation suggests FAISS may offer more reliable distance-based scoring and novelty inference than ScaNN, whose distribution remains narrow and less distinguishable across query types.

\section{Conclusion}

We benchmarked FAISS and ScaNN for high-throughput similarity search on dense gene fragment embeddings from metagenomic data. Across comprehensive evaluations, FAISS consistently outperformed ScaNN in indexing speed, query latency, retrieval accuracy, and novelty detection performance. In particular, FAISS's broader and more discriminative distance distributions enabled more interpretable and effective retrieval—especially for distinguishing in-domain and out-of-domain sequences.

While ScaNN offered faster inference in certain configurations and lower CPU usage during search, its overall accuracy and sensitivity to taxonomic novelty were limited. Parameter sweeps and SHAP analysis confirmed that ScaNN tuning is most useful for optimizing runtime rather than improving accuracy.

To contextualize our findings, we also compared both embedding-based methods to MMseqs2, a traditional alignment-based tool. Although MMseqs2 is widely used, it performed poorly on short 400 bp fragments—achieving far lower accuracy and higher latency—highlighting the growing advantage of learned representations in metagenomic applications.

It is important to note that all evaluations were conducted on CPUs due to infrastructure limitations. Both FAISS and ScaNN support GPU acceleration, and future work should investigate how GPU-based execution affects indexing scalability, query latency, and memory usage at larger scales.

Overall, our results support the use of embedding-based retrieval frameworks like FAISS for scalable and biologically meaningful similarity search in large genomic datasets.


\bibliography{refs}
\bibliographystyle{icml2025}




\end{document}